\documentclass{article}
\usepackage[utf8]{inputenc}
\usepackage{authblk}
\usepackage{indentfirst}
\usepackage{hyperref}
\usepackage{longtable}
\usepackage{tabularray}
\usepackage{setspace}
\usepackage[margin=1in]{geometry}
\usepackage{graphicx}
\usepackage{subcaption}
\usepackage{amsmath}
\usepackage{lineno}
\usepackage{multirow}
\usepackage{color,soul}
\usepackage[version=3]{mhchem}

\usepackage[biblabel]{cite}

\title{Exploration of oxyfluoride frameworks as Na-ion cathodes}

\author[1]{Debolina Deb}
\author[1,*]{Gopalakrishnan Sai Gautam}

\affil[1]{Department of Materials Engineering, Indian Institute of Science, Bengaluru, 560012, India}
\affil[*]{Email: \href{mailto:saigautamg@iisc.ac.in}{saigautamg@iisc.ac.in}}

\date{}

\begin{document}

\maketitle

\begin{abstract}
Na-ion batteries (NIBs) are increasingly looked at as a viable alternative to Li-ion batteries due to the abundance, low cost, and thermal stability of Na-based systems. To improve the practical utilization of NIBs in applications, it is important to boost the energy and power densities of the electrodes being used, via discovery of novel candidate materials. Thus, we explore the chemical space of transition metal containing oxyfluorides (TMOFs) that adopt the perovskite structure as possible NIB electrodes. Our choice of the perovskite structure is motivated by the `large' cationic tunnels that can accommodate Na$^+$, while the chemistry of TMOFs is motivated by the high electronegativity and inductive effect of F$^-$, which can possibly lead to higher voltages. We use density functional theory based calculations to estimate the ground state polymorphs, average Na (de)intercalation voltages, thermodynamic stabilities and Na$^+$ mobility on two distinct sets of compositions: the F-rich Na$_{x}$MOF$_{2}$, and the O-rich Na$_{1+x}$MO$_{2}$F where $x$ = 0--1 and M~=~Ti, V, Cr, Mn, Fe, Co, or Ni. Upon identifying the ground state polymorphs in the charged compositions (i.e., MOF$_2$ and NaMO$_2$F), we show that F-rich perovskites exhibit higher average voltages compared to O-rich perovskites. Also, we find six stable/metastable perovskites in the F-rich space, while all O-rich perovskites (except NaTiO$_2$F) are unstable. Finally, our Na-ion mobility calculations indicate that TiOF$_{2}$-NaTiOF$_2$, VOF$_{2}$-NaVOF$_2$, CrOF$_{2}$, and NaMnOF$_{2}$ can be promising compositions for experimental exploration as NIB cathodes, primarily if used in a strained electrode configuration and/or thin film batteries. Our computational approach and findings provide insights into developing practical NIBs involving fluorine-containing intercalation frameworks.
\end{abstract}


\section{Introduction}

Na-ion battery (NIB) technology is a key contributor in reducing the extensive dependence on Li-ion batteries (LIBs) to fulfil the ever-increasing energy demands.\cite{chayambuka2020li,vaalma2018cost,zhang2019comprehensive,xiao2021recent,kim2012electrode} As a technology, NIBs have come a long way with notable applications in both electric vehicles and stationary energy storage.\cite{rudola2023opportunities,barker2017storage,li2024secondary,hirsh2020sodium}  Nevertheless, the practical utility of NIBs can be further enhanced with the development of novel high energy and power density electrode materials. While layered transition-metal oxides (TMOs) are the state-of-the-art NIB positive electrodes (cathodes),\cite{wei2021research} the structural instabilities of layered compounds at their fully desodiated states and detrimental phase transitions have directed research towards polyanionic cathode frameworks.\cite{kim2020direct,deb2022critical,mariyappan2018will} Some of the most explored polyanionic frameworks, such as the sodium superionic conductors (NaSICONs), alluaudites, olivines, and pyro/fluoro-phosphates display a wide range of electrochemical performance and good structural stability, with low gravimetric capacity being a common impediment.\cite{deb2022critical} Thus, an ideal NIB cathode must be able to (de)intercalate the large Na$^+$ at high rates, without compromising on structural stability, and deliver a large capacity for achieving both high energy and power densities. An ideal NIB negative electrode (anode) has similar requirements as the ideal cathode as well.

Oxide perovskites, which have a general formula of ABO$_{3}$, where A and B are cations, have been explored for several non-energy-storage applications, due to their structural stability and compositional flexibility.\cite{tian2022perovskite,ye2023design,assirey2019perovskite,fakharuddin2022perovskite,sai2020exploring} Importantly, perovskites are suitable structures for accommodating Na$^+$ because of their rigid open structures with large voids.\cite{evans2020perovskite} Additionally, incorporation of fluorides in cathode frameworks often leads to improved energy densities, since the higher electronegativity of F$^-$ typically leads to a higher (de)intercalation voltage via the induction effect.\cite{wang2024fluorination,manthiram1989lithium,padhi1997phospho} Indeed, many of the best-performing polyanionic NIB cathodes contain fluorine.\cite{yan2019higher,nguyen2020combined,akhtar2023novel} Thus, fixing the A cation in a perovskite as Na$^+$, the B cation to be a redox-active 3$d$ transition metal (TM), and the anions being a mixture of both O and F, yields a class of perovskite-based TM oxyflouride (TMOFs) compositions as potential NIB cathodes (or anodes). 

So far, perovskite TMOFs are a largely unexplored class of battery cathodes (or anodes), primarily due to synthesis difficulties from highly stable fluoride precursors.\cite{deng2017transition} Indeed, only a few TMOFs, including, TiOF$_{2}$ (space group: \textit{Pm}$\overline{3}$\textit{m}),\cite{reddy2006metal} VO$_2$F (\textit{R}$\overline{3}$\textit{c}),\cite{kuhn2020redox,wang2018structural} and NbO$_{2}$F (\textit{Pm}$\overline{3}$\textit{m}),\cite{reddy2006metal} have been investigated as LIB cathodes. Additionally, Li$_{2}$MO$_{2}$F with M across the 3$d$ series\cite{xu2021li2nio2f} and Na$_{2}$MnO$_{2}$F\cite{shirazi2022rich} have been reported to exhibit a disordered rocksalt, and not a perovskite-based structure. Also, most of the oxyfluoride structures that have been reported have undergone either amorphization or an irreversible structural transition during electrochemical cycling.\cite{perez2015vo,chen2016lithiation,astrova2023titanium,kitajou2018electrochemical} Although Li-ion mobility is not hindered in both disordered rocksalt,\cite{clement2020cation} and amorphized,\cite{ye2018amorphization} oxyfluorides, studies have not analysed Na-ion mobility in similar frameworks. Notably, the rutile-FeOF (\textit{P}4$_{2}$/\textit{mnm}) structure was tested as a NIB cathode and showed a reversible transition to cubic-Na$_{x}$FeOF.\cite{kitajou2018electrochemical} However, this FeOF$\leftrightarrow$Na$_x$FeOF transition was accompanied by significant hysteresis in the corresponding voltage-capacity profiles, with possible contributions from electrolyte decomposition and/or other side reactions.\cite{kitajou2018electrochemical} Importantly, the chemical class TMOFs has not been systematically explored, either computationally or experimentally, as NIB cathodes, so far.

Here, we present a systematic density functional theory (DFT)-based computational exploration of perovskite-based TMOF compositions as potential NIB cathodes (or anodes). Specifically, we explore the chemical compositions of oxygen-rich  (NaMO$_2$F$\leftrightarrow$Na$_2$MO$_2$F) and fluorine-rich (MOF$_2 \leftrightarrow$NaMOF$_2$) perovskites, where M = Ti, V, Cr, Mn, Fe, Co, or Ni. For both O-rich and F-rich compositions, we examine possible crystalline structures of the general perovskite framework. Importantly, we have evaluated the ground state Na-vacancy configurations, average Na intercalation voltages, and the 0~K thermodynamic stabilities in both O-rich and F-rich TMOFs, followed by an evaluation of the Na-ion mobility in a subset of candidate compounds. Besides shedding light on the overall trends in voltages and stabilities, we also identify a few promising compositions, namely, TiOF$_{2}$-NaTiOF$_{2}$, VOF$_{2}$-NaVOF$_{2}$, CrOF$_{2}$ and NaMnOF$_{2}$, as candidate NIB electrodes which can be relevant for subsequent experimental validation, primarily in strained configurations. We hope that our study opens up the novel oxyfluoride chemical space for battery cathode applications and beyond.

\section{Methods and Workflow}
\subsection{Structure identification}
To explore the TMOF chemical space, we used the charged O-rich (i.e., NaMO$_2$F) and F-rich (MOF$_2$) compositions as the initial cases for structure generation for all TMs. Note that both the charged compositions correspond to the TM being in a +4 oxidation state, while the corresponding discharged compositions (Na$_2$MO$_2$F and NaMOF$_2$) reflect the TM in a +3 oxidation state. To identify the relevant space group/polymorph, we first searched the inorganic crystal structure database (ICSD\cite{hellenbrandt2004inorganic}) for experimental structures with the NaMO$_2$F and MOF$_2$ compositions, where we found only TiOF$_{2}$ (ICSD collection code 160661; \textit{Pm}$\overline{3}$\textit{m}), VO$_{2}$F (ICSD collection code 142594; \textit{R}$\overline{3}$\textit{c}) and LiVO$_{2}$F (ICSD collection code 142596; \textit{R}$\overline{3}$\textit{c}). Thus, we used the TiOF$_2$ structure from the ICSD as the starting configuration for all calculations involving cubic-TiOF$_2$, VO$_{2}$F for calculations of rhombohedral-MOF$_{2}$, and LiVO$_2$F for calculations of rhombohedral-NaMO$_2$F.   

\begin{figure*}
    \centering
    \includegraphics[width=\linewidth]{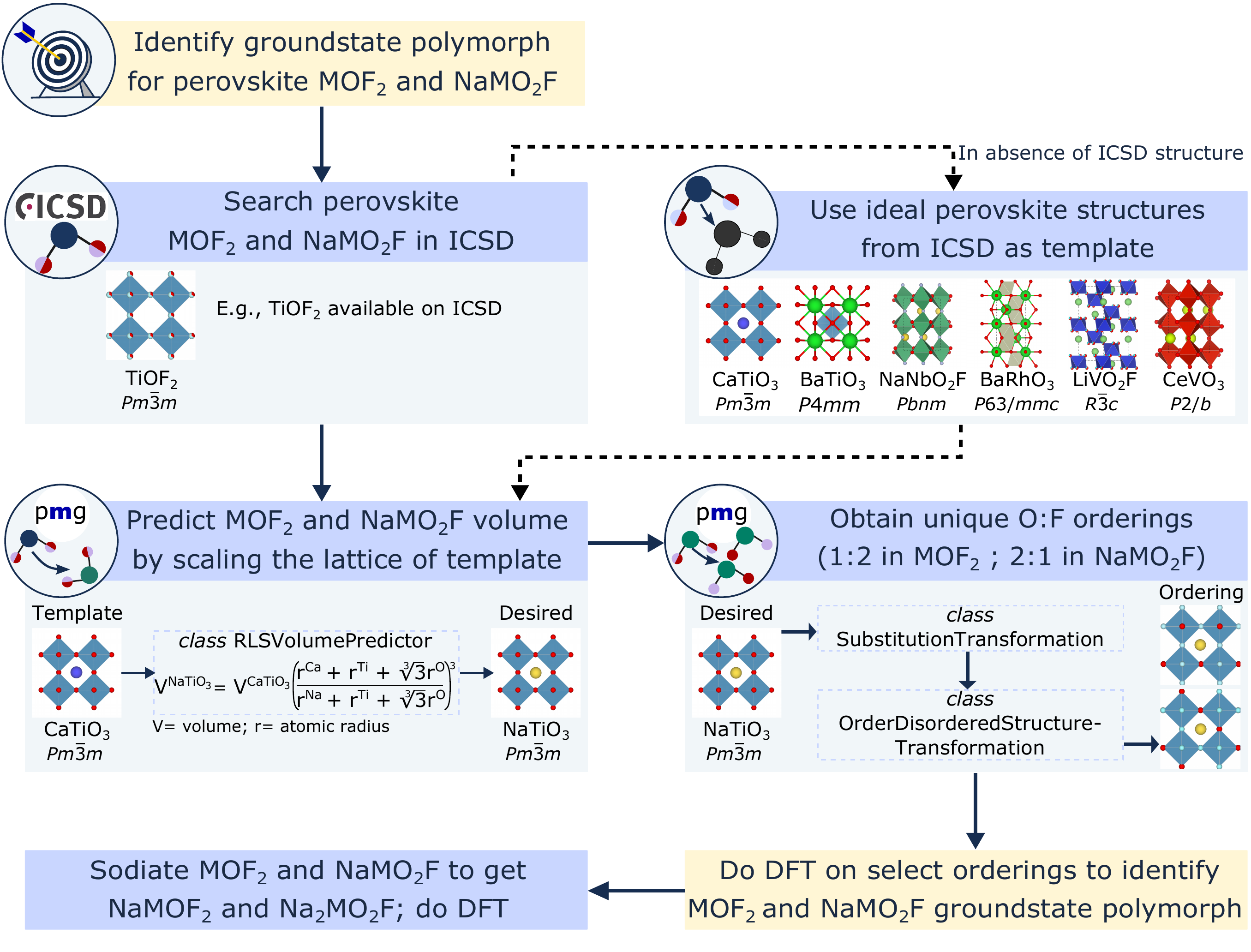}
    \caption{Workflow to obtain ground state polymorph for desodiated F-rich MOF$_{2}$ and O-rich NaMO$_{2}$F perovskites, where M = Ti, V, Cr, Mn, Fe, Co, or Ni. Notations `pmg' and `ICSD' refer to the pymatgen package and the inorganic crystal structure database.}
    \label{fig:workflow}
\end{figure*}

Given the absence of ICSD structures for other TMOFs, we theoretically generated possible structures for both charged compositions, using the workflow displayed in \textbf{Figure~{\ref{fig:workflow}}}. Similar to the procedure used in a previous study\cite{sai2020exploring}, we used experimental template structures among six different space groups that are commonly adopted by perovskite compositions to generate six possible theoretical structures for each composition. Specifically, we used CaTiO$_3$, BaTiO$_3$, NaNbO$_2$F, BaRhO$_3$, LiVO$_2$F, and CeVO$_3$ as templates for the \textit{Pm}$\overline{3}$\textit{m}, \textit{P}4\textit{mm}, \textit{Pbnm}, \textit{P}63/\textit{mmc}, \textit{R}$\overline{3}$\textit{c}, and \textit{P}2/\textit{b} space groups, respectively. We chose Ba and Ca containing structures as templates due to the similarity in ionic radii of Ba$^{2+}$ and Ca$^{2+}$ to Na$^+$. CeVO$_3$ was the only reasonable monoclinically-distorted perovskite template we could find. For rhombohedral perovskites, the presence of VO$_2$F and LiVO$_2$F experimental structures provided us both an oxyfluoride template along with possible Na sites, as Li can be substituted with Na,\cite{heubner2017situ} motivating our use of LiVO$_2$F as the template.\cite{kuhn2020redox} As far the orthorhombic perovskite, NaNbO$_2$F is an oxyfluoride and contains Na, and hence was the obvious choice as a template. Note that we used the TiOF$_2$ structure as the \textit{Pm}$\overline{3}$\textit{m} template for other MOF$_2$ compositions, while we used the CaTiO$_3$ structure as the \textit{Pm}$\overline{3}$\textit{m} template for NaMO$_2$F compositions. In addition, we used VO$_2$F structure as the \textit{R}$\overline{3}$\textit{c} template for all rhombohedral-MOF$_2$ compositions and LiVO$_2$F structure as the \textit{R}$\overline{3}$\textit{c} template for all rhombohedral-NaMO$_2$F compositions.

From each template structure, we performed chemical substitution (i.e., replace Ca/Ba/Li/Ce with Na, and the remaining cation with a 3$d$ TM), to result in a NaMO$_3$ composition. Subsequenty, we used the RLSVolumePredictor\cite{chu2018predicting} class of the pymatgen package to scale the lattice parameters of the template structure to values that better represent a NaMO$_3$ perovskite composition. Upon lattice scaling, we introduced F, based on a O:F ratio of 2:1 in O-rich perovskites, and 1:2 in F-rich perovskites, by inducing disorder within the anionic sublattice using the SubstitutionTransformation class of pymatgen. Note that in F-rich perovskites, we removed the Na before the lattice scaling step. Finally, we enumerated symmetrically distinct O-F arrangements for all distinct template space groups in both the NaMO$_2$F and MOF$_2$ compositions, using the OrderDisorderedStructureTransformation class of pymatgen, and performed DFT calculations to determine the respective ground state configurations. During enumerations, we took a maximum of 16 structures that exhibited the lowest electrostatic energy, calculated using the Ewald summation technique,\cite{ewald1921berechnung} to minimize computational expense. In the case of \textit{P4mm} and \textit{R}$\overline{3}$\textit{c} perovskites (both MOF$_2$ and NaMOF$_2$), we obtained a total of only five and three symmetrically distinct configurations upon enumeration and all configurations were considered for DFT calculations. In the case of \textit{Pm}$\overline{3}$\textit{m}, \textit{Pbnm}, \textit{P}63/\textit{mmc}, and \textit{P}2/\textit{b} space groups, we obtained a total of 22, 40, 55, and 48 symmetrically distinct configurations, respectively, out of which we chose the 16 lowest electrostatic energy configurations for each space group (for both MOF$_2$ and NaMOF$_2$).

\begin{figure*}
    \centering
    \includegraphics[width=\linewidth]{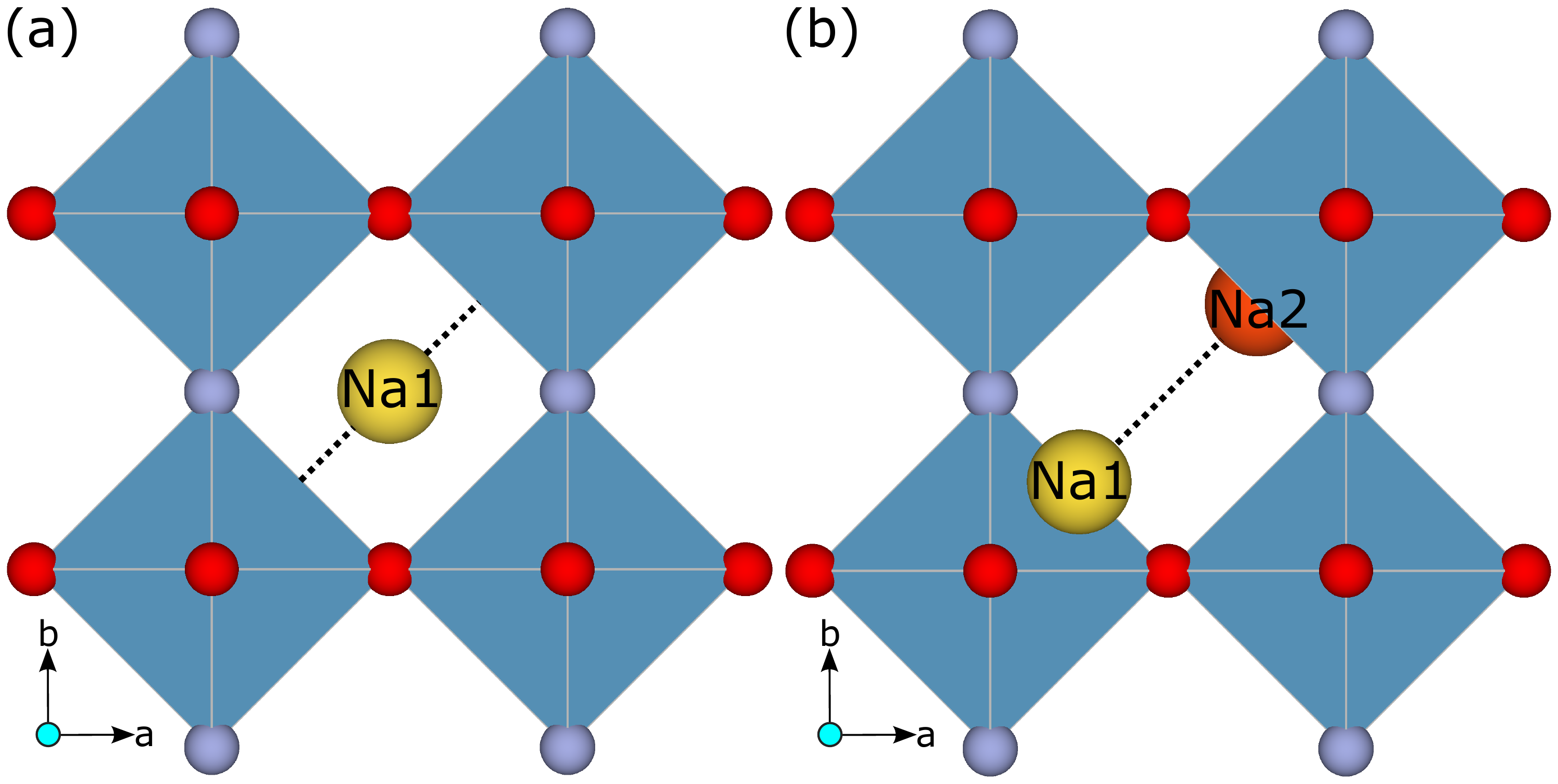}
    \caption{(a) NaTiO$_2$F with the initial Na atom (denoted by Na1) at the centre of the cube (i.e., fractional coordinates of (0.5, 0.5, 0.5)). (b) Displacement of the Na1 atom to (0.25, 0.25, 0.25) and subsequent occupation of the second Na atom (denoted by Na2) at (0.75, 0.75, 0.75). Blue polyhedra in both panels denote TiO$_4$F$_2$ octahedra. O and F are represented by red and purple spheres, respectively. Body diagonals within the cubic structures are indicated by the dotted black lines.}
    \label{fig:na2}
\end{figure*}

Once the ground state polymorph of each desodiated NaMO$_2$F and MOF$_2$ composition was determined, we added Na to the DFT-relaxed charged ground state structures to obtain the corresponding discharged (or sodiated) configurations, i.e., Na$_2$MO$_2$F and NaMOF$_2$. For NaMOF$_{2}$, we initialised the Na ions on the sites occupied by the A-cation in the corresponding template perovskite structure. Given that the NaMOF$_2$ perovskite only has one distinct Na (or A cation) site, we created a second Na site by displacing the existing Na ion to minimize electrostatic repulsion between the two Na ions, as displayed in \textbf{Figure~{\ref{fig:na2}}}. For example, in the case of 
\textit{Pm}$\overline{3}$\textit{m} NaTiO$_2$F, we displaced the existing Na from the centre of the cube (i.e., fractional coordinates of (0.5, 0.5, 0.5)) along the body diagonal to a new site of coordinates (0.25, 0.25, 0.25). Subsequently, we initialised the second Na atom at the coordinates of (0.75, 0.75, 0.75), to minimize electrostatic repulsions between the two Na. The introduction of additional Na sites in other perovskite structures is described in the supporting information (SI), along with a schematic in \textbf{Figure~S1}. 

\subsection{Computational details}
We used the Vienna ab initio simulation package (VASP\cite{kresse1996efficiency,kresse1996efficient}) for all spin-polarized DFT calculations. We utilised the projector augmented-wave (PAW\cite{kresse1999ultrasoft,blochl1994projector}) potentials similar to our previous work,\cite{gautam2018evaluating,long2020evaluating,swathilakshmi2023performance} with the list of PAW potentials used in this work compiled in \textbf{Table~S1}. To account for the electronic exchange and correlation, we employed the Hubbard $U$ corrected\cite{dudarev1998electron,anisimov1991band} strongly constrained and appropriately normed (i.e., SCAN+$U$)\cite{sun2015strongly,gautam2018evaluating,long2020evaluating} functional. We utilised $U$ values that were obtained for TMOs in our work\cite{gautam2018evaluating,long2020evaluating}, since they gave the best agreement between calculated and experimental average voltages in Li-based TMOFs (see \textbf{Table~S2}). We expanded the one-electron wavefunctions using a plane wave basis set, with a 520~eV kinetic energy cutoff, and used a Gaussian smearing of width 0.05~eV to integrate the Fermi surface. We sampled the irreducible Brillouin zone with a $\Gamma$-centered Monkhorst-pack\cite{monkhorst1976special} $k$-mesh with a density of at least 32 $\textit{k}$-points per \AA$^{-1}$ (i.e., a minimum sampling of 32 subdivisions along each reciprocal lattice vector). For the total energies and atomic forces, we set the convergence criterion to be 0.01 meV and $|$0.03$|$ eV/\AA$^{-1}$, respectively. To reduce computational complexity, we initialised all 3$d$ TMs in their corresponding high-spin ferromagnetic configurations. For all structures, we relaxed the cell volume, cell shape, and ionic positions without preserving any symmetry. Where possible, we have followed a colour-blind friendly colour scheme in our plots.\cite{Hunter:2007}

\subsection{Ab initio thermodynamics}
For evaluating the thermodynamic stability of the TMOFs considered, we constructed the 0~K convex hull of the corresponding quaternary (i.e., Na-TM-O-F) chemical spaces using the pymatgen package. Specifically, we collected experimentally-reported structures of individual elements (Na, TM, O, and F), binaries (Na-O, Na-F, TM-O, and TM-F), ternaries (Na-O-F, TM-O-F, Na-TM-O, Na-TM-F), and quaternaries (Na-TM-O-F) from the ICSD, and subsequently calculated their total energies using DFT. Note that we only considered ICSD structures that were fully ordered, i.e., each lattice site in a structure exhibits an integer occupation of a given species. Also, for individual elements, Na-O and Na-F binaries, and Na-O-F ternaries we used only the SCAN functional for treating the electronic exchange and correlation, while for the other structures we used the SCAN+$U$ functional. Since we have utilised only DFT-calculated total energies to construct the 0~K convex hull, our phase diagrams do not include the $p-V$ contributions. 

Importantly, any stable entity on the 0~K convex hull will have a energy above convex hull (E$^{hull}$) as 0~meV/atom, while any metastable/unstable entity will have E$^{hull} >$0.\cite{gautam2019theoretical} Given that compounds that are metastable at 0~K can be stabilised under different experimental conditions, we used a synthesizeability threshold of E$^{hull} \leq$100~meV/atom.\cite{balachandran2018predictions} This implies that compounds with a E$^{hull} \leq$100~meV/atom may be synthesizeable under higher temperatures/pressures, and can be considered metastable, while compounds with E$^{hull} >$100~meV/atom are unlikely to be synthesizeable and can be considered to be unstable. All calculated phase diagrams (except for the Na-Ti-O-F quaternary) are compiled in \textbf{Figure~S3}, while the list of stable/unstable compounds are compiled in \textbf{Table~S4}.

The average voltage for Na (de)intercalation in TMOFs is evaluated using DFT-based total energies from the well-known Nernst equation.\cite{aydinol1997ab} Considering a Na (de)intercalation reaction of the form, Na$_x$TMOF + $\Delta x$Na $\leftrightarrow$ Na$_{x+\Delta x}$TMOF, we can approximate the Gibbs energy change ($\Delta G$) associated with the (de)intercalation process using \textbf{Equation~{\ref{eqn:voltage}}}, which neglects entropic and $p-V$ contributions. Note that the $E$ terms in \textbf{Equation~{\ref{eqn:voltage}}} are DFT-calculated, with Na$_x$TMOF and Na$_{x+\Delta x}$TMOF described with SCAN+$U$ and metallic Na described with SCAN in its body-centered-cubic ground state. $F$ is Faraday's constant.

\begin{equation}
\label{eqn:voltage}
    {\langle V \rangle} = \frac{\Delta G}{\Delta x~F} \approx -\frac{E(\mathrm{Na}_{x+\Delta x}\mathrm{TMOF})-[E(\mathrm{Na}_x\mathrm{TMOF})+\Delta x~E(\mathrm{Na})]}{\Delta x~F}
\end{equation}

\subsection{Kinetics}
To estimate the ionic mobility of Na$^+$ in select TMOF frameworks, we utilized DFT-based nudged elastic band (NEB\cite{sheppard2008optimization,henkelman2002methods}) calculations to estimate the migration barrier (E$_m$) associated with Na$^+$ motion. For all structures, we considered a vacancy-mediated Na$^+$ migration along the A-site `tunnel' of the perovskite framework, and calculated E$_m$ either at the charged or the discharged sodium concentration limits. Upon introducing a Na-vacancy and fully relaxing the endpoint configurations, we interpolated five images across the endpoints to initialise the minimum energy path (MEP). 

Spring forces of 5~eV/{\AA} were introduced between the images, and we considered the NEB calculation converged when the total energy of each image and the perpendicular component of the force between each image dropped below 0.01~meV and $|$0.05$|$~eV/{\AA}, respectively. For all NEB calculations, we used supercells with lattice parameters $\geq$8~{\AA} to avoid spurious interactions of the migrating Na with its periodic images. We used the Perdew-Burke-Ernzerhof (PBE\cite{perdew1996generalized}) parameterization of the generalized gradient approximation (GGA) to describe the exchange-correlation in our NEB calculations instead of SCAN, since GGA provides accurate qualitative trends at lower computational cost and with fewer convergence difficulties.\cite{devi2022effect} All computed MEPs are compiled in \textbf{Figure~S4}.

\section{Results}
\subsection{Ground state polymorphs and average voltages}
\begin{figure*}
    \centering
    \includegraphics[width=\linewidth]{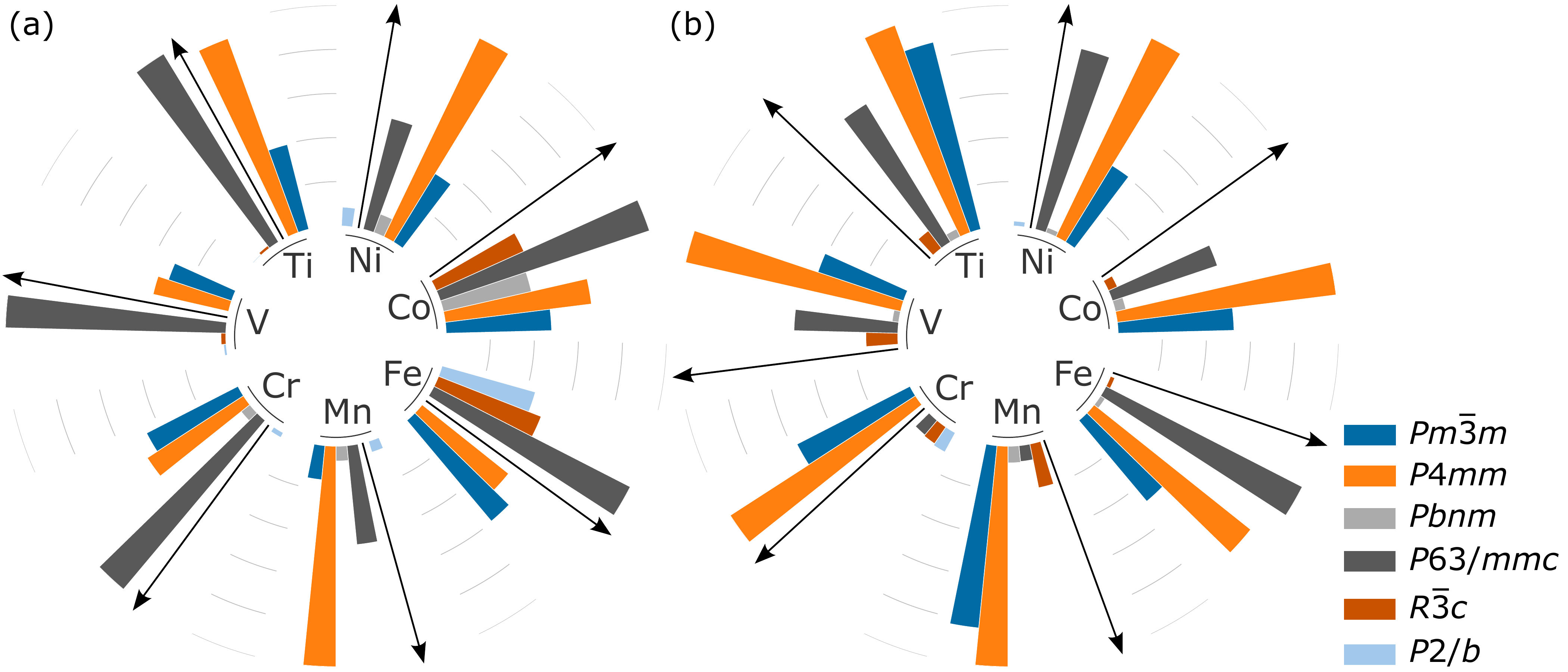}
    \caption{Percentage normalised relative energies of all polymorphs considered with respect to the corresponding ground states for (a) F-rich MOF$_{2}$ and (b) O-rich NaMO$_{2}$F. All ground state polymorphs are indicated by black arrows. Each concentric ring on the radars represent a percentage step of 20\%.}
    \label{fig:groundstate}
\end{figure*}

The ground state polymorph for each desodiated F-rich MOF$_{2}$ and O-rich NaMO$_{2}$F are represented by the black arrows in panels a and b of \textbf{Figure~{\ref{fig:groundstate}}}. The percentage normalised differences in energies of the other polymorphs considered, relative to the ground state, are plotted as bars in \textbf{Figure~{\ref{fig:groundstate}}}. Specifically, we have plotted the percentage differences, calculated as $\frac{E(\mathrm{polymorph})-E(\mathrm{ground state})}{E(\mathrm{highest-energy-polymorph})-E(\mathrm{ground state})}\times 100$, where each concentric ring on the radars represent percentage steps of 20\%. Thus, the ground state and the highest energy polymorph represent 0\% and 100\%, respectively, on the radars of \textbf{Figure~{\ref{fig:groundstate}}} for each composition. Notably, the ground state polymorphs of the MOF$_2$ compositions include \textit{Pbnm} (for Ti, V, Fe), \textit{R}$\overline{3}$\textit{c} (Cr, Mn, Ni), and \textit{P2/b} (Co), while for the NaMO$_2$F compositions are \textit{P2/b} (Ti, V, Mn, Fe, Co), \textit{Pbnm} (Cr), and \textit{R}$\overline{3}$\textit{c} (Ni). We have compiled the percentage normalised relative energies for all perovskite polymorphs considered in \textbf{Table~S3} and provided schematics of the desodiated ground states and their corresponding sodiated structures in \textbf{Figure~S2}.    

\begin{figure*}
    \centering
    \includegraphics[width=\linewidth]{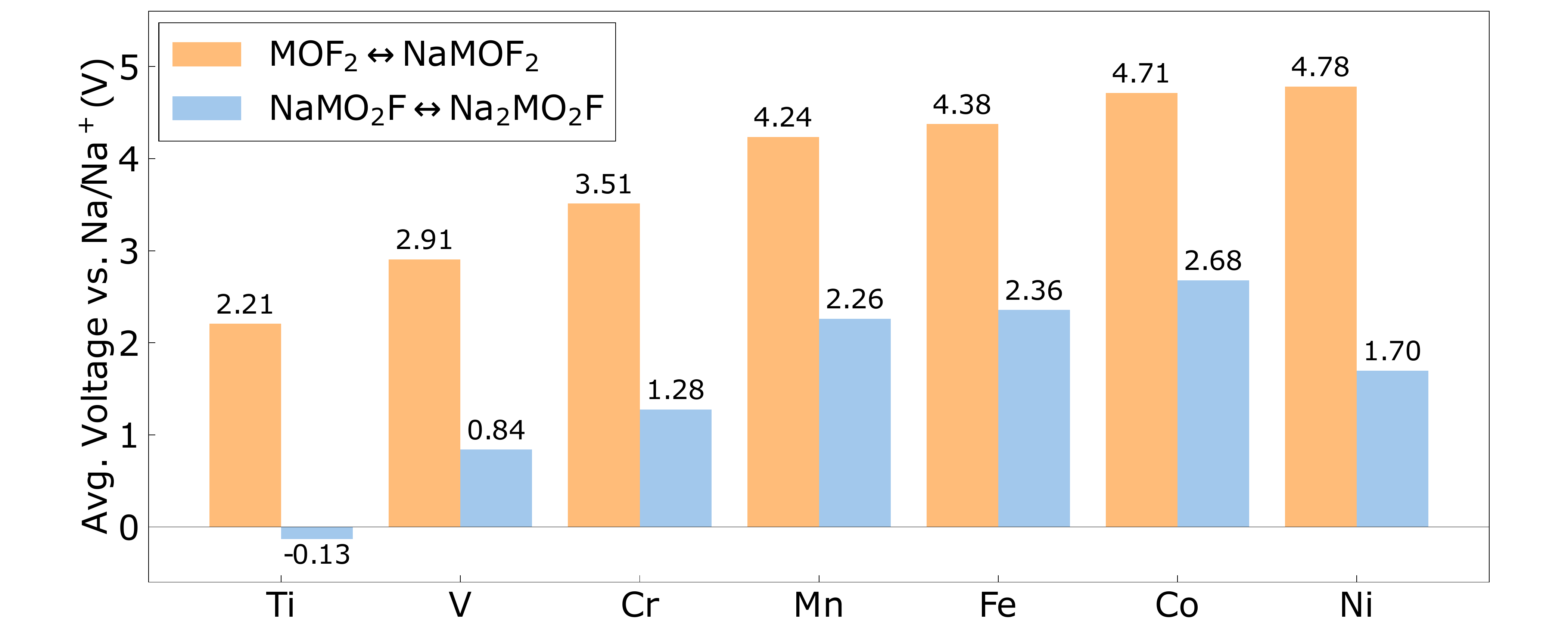}
    \caption{Calculated average Na (de)intercalation voltages (in V), versus Na/Na$^+$, in F-rich (orange bars) and O-rich (blue bars) TMOFs considered.}
    \label{fig:voltage}
\end{figure*}

\textbf{Figure~{\ref{fig:voltage}}} depicts the calculated average voltages for Na (de)intercalation, versus Na/Na$^{+}$, into the ground state polymorphs of F-rich MOF$_{2}$ (orange bars) and O-rich NaMO$_{2}$F (blue bars). The extent of Na (de)intercalation considered in both F-rich and O-rich perovkistes are one Na per f.u., corresponding to MOF$_{2}$ $\leftrightarrow$ NaMOF$_{2}$ and NaMO$_{2}$F$\leftrightarrow$ Na$_{2}$MO$_{2}$F, respectively. Expectedly, we find the F-rich perovskites to exhibit consistently higher average voltages than the corresponding O-rich perovskites, which can be attributed to the greater inductive effect of F$^-$ compared to O$^{2-}$.\cite{wang2024fluorination} Indeed, fluorine's inductive effect causes an increase in average voltage of $\geq$2~V for all TM (except Mn at a 1.98~V increase), with the increase in Ni being the highest at 3.08~V.

In both the F-rich and O-rich perovskites, there is a monotonic increase in voltages along the 3$d$ series, with the values increasing from 2.21~V (in Ti) to 4.78~V (Ni) in F-rich, and $-$0.13~V (Ti) to 2.68~V (Co) in O-rich. The monotonic trends in voltages can be largely attributed to the corresponding trends in standard reduction potentials of the TMs.\cite{vanysek2000electrochemical} The dip in voltage from Co to Ni in O-rich perovskites can be primarily attributed to cooperative Jahn-Teller distortion in the Ni-perovskite, which results in a larger deviations in lattice parameters (see compiled $b/a$ and $c/a$ ratios in Mn- and Ni-perovskites in \textbf{Table~S5}). Interestingly, the average intercalation voltage in the O-rich Ti-perovskite exhibits a negative value ($-$0.13~V), indicating non-spontaneity of Na-intercalation in this system. This is because the intercalated Na$_2$TiO$_2$F is thermodynamically unstable, with Na-metal being one of the decomposition products (see \textbf{Figure~{\ref{fig:Ti-stability}}}).

Given the electrolyte stability windows of liquid electrolytes in NIBs typically span up to 4.8~V vs. Na/Na$^+$,\cite{karahan2023recent,wu2016highly} we find all perovskites considered in this work to be suitable as NIB electrodes. The low average voltages of several O-rich perovskites, including V, Cr, Mn, and Fe ($<$2.5~V), and TiOF$_2$, make these systems more suitable as negative electrodes (anodes) than cathodes in a NIB. Thus, based on the voltage data alone, we find the Mn-, Fe-, Co-, and Ni-based F-rich perovskites to be the most promising as NIB cathodes. However, practical deployment of candidate electrode materials in NIBs will be highly dependent on their synthesizability (i.e., thermodynamic stability) and their rate performance (i.e., Na-ion mobility). 

\subsection{Thermodynamic stability}
\begin{figure*}
    \centering
    \includegraphics[width=\linewidth]{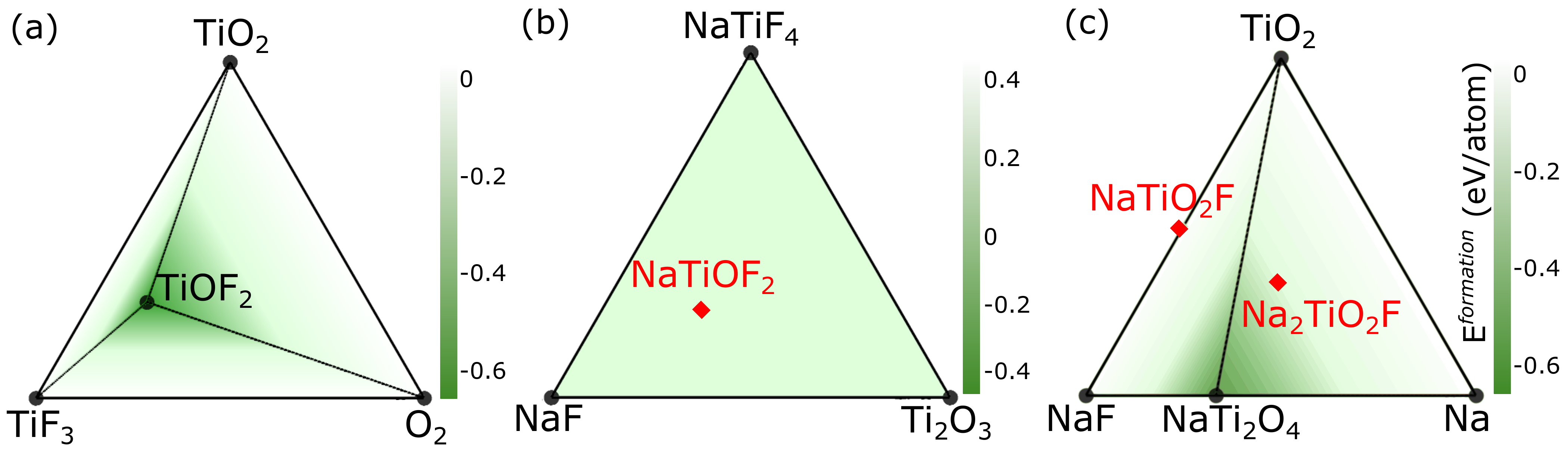}
    \caption{Ternary projections of Na-Ti-O-F phase diagram to visualise (a) TiOF$_{2}$, (b) NaTiOF$_{2}$, and (c) NaTiO$_{2}$F and Na$_{2}$TiO$_{2}$F. In each panel, the green-to-white background represents E$^{formation}$, while the red diamonds indicate meta/instability (E$^{hull} >$0). Black circles indicate stable compositions and black lines are tie-lines.} 
    \label{fig:Ti-stability}
\end{figure*}

Upon construction of the quaternary 0~K Na-TM-O-F convex hulls, we plotted pseudo-ternary slices (or projections) of the quaternary phase diagram for each TM for ease of visualization. For instance, ternary projections of the Na-Ti-O-F system is displayed in \textbf{Figure~{\ref{fig:Ti-stability}}}, while the ternary projections for the remaining TM systems are compiled in \textbf{Figure~S3}. The background colors in all panels (shades of green) of \textbf{Figure~{\ref{fig:Ti-stability}}} indicate the energy of formation (E$^{formation}$), calculated with respect to the terminating compositions of the ternary projections. Stable compounds within the ternary projections are indicated by black circles. Metastable/unstable compounds are indicated by red diamonds. For each metastable/unstable compound among the TMOFs considered, the set of decomposition products (i.e., stable compounds that a metastable/unstable compound is thermodynamically driven to decomposes into) is compiled in \textbf{Table~S6}.

For visualising the TiOF$_2$ composition in the Ti-quaternary, we used a ternary projection terminated by TiO$_2$, TiF$_3$, and O$_2$ (see \textbf{Figure~{\ref{fig:Ti-stability}}a}). Similarly, for visualizing the NaTiOF$_2$, we used the NaTiF$_4$-NaF-Ti$_2$O$_3$ ternary projection (\textbf{Figure~{\ref{fig:Ti-stability}}b}). Both NaTiO$_2$F and Na$_2$TiO$_2$F can be captured within the TiO$_2$-NaF-Na projection \textbf{Figure~{\ref{fig:Ti-stability}}c}. Importantly, \textbf{Figure~{\ref{fig:Ti-stability}}} indicates that TiOF$_2$ is thermodynamically stable (E$^{hull} =$0~meV/atom), NaTiOF$_2$, and NaTiO$_2$F are metastable with E$^{hull}$ of 46, and 53~meV/atom, respectively, which are below the 100~meV/atom threshold, and Na$_2$TiO$_2$F is unstable with E$^{hull}$ of 187~meV/atom. Notably, Na-metal is one of the decomposition products for the unstable Na$_2$TiO$_2$F (\textbf{Figure~{\ref{fig:Ti-stability}}c}), which explains the calculated negative intercalation voltage for the NaTiO$_2$F $\leftrightarrow$ Na$_2$TiO$_2$F reaction (\textbf{Figure~{\ref{fig:voltage}}}).   

\begin{figure*}
    \centering
    \includegraphics[width=\linewidth]{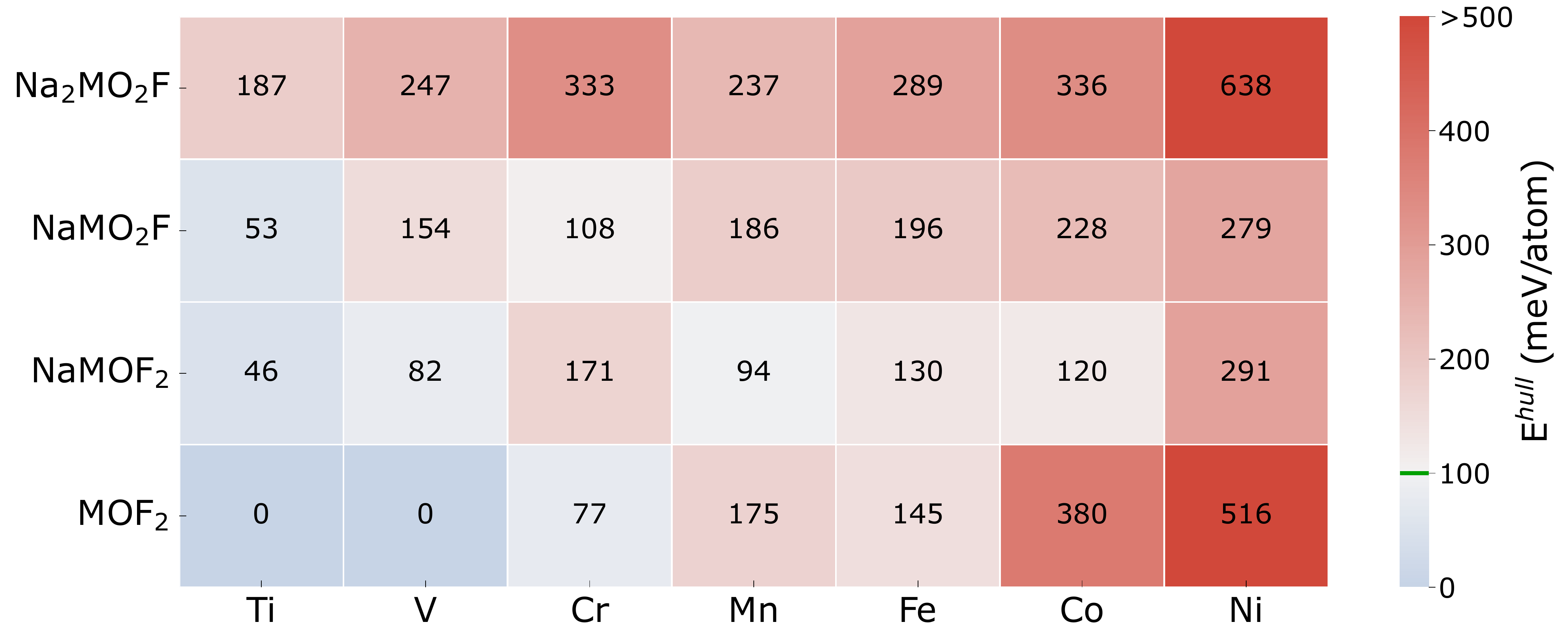}
    \caption{Calculated E$^{hull}$) for charged and discharged F-rich TMOFs (bottom two rows) and O-rich TMOFs (top two rows). Each column represents a 3$d$ TM, while the E$^{hull}$ for each compound is indicated using text annotations within each square. The green line on the legend bar indicates the 100~meV/atom synthesizability threshold considered in this work.}
    \label{fig:ehull}
\end{figure*}

The heatmap depicted of \textbf{Figure~{\ref{fig:ehull}}} compiles the E$^{hull}$ data of all charged and discharged O-rich and F-rich perovskites considered. Blue squares indicate stable/metastable compounds, while red squares indicate unstable compounds. The text annotations within each square represents the E$^{hull}$ in meV/atom for the corresponding compound. Significantly, we find only TiOF$_{2}$ and VOF$_{2}$ to be thermodynamically stable (i.e., E$_{hull} =$0~meV/atom) among all the TMOFs considered. This is in agreement with experimental reports that have synthesized TiOF$_2$\cite{reddy2006metal} and VOF$_2$.\cite{koksbang1996lithium} All charged and discharged compositions of Fe-, Co-, and Ni-based TMOFs are unstable, with E$^{hull}$ greater than the synthesizability threshold of 100~meV/atom, citing the high unsuitability of such compositions as NIB electrodes. Moreover, all O-rich perovskites, except NaTiO$_2$F, exhibit E$^{hull} >$100~meV/atom, highlighting their high instabilities.

While it is good for an electrode to have thermodynamically stable charged and discharged states to avoid any irreversible decomposition or conversion reactions during an electrochemical cycle, topotactic (de)intercalation is often possible with metastable charged and discharged states as well.\cite{hannah2018balance,malik2013critical,tekliye2022exploration,amatucci1996cobalt} Thus, compositions that lie within the E$^{hull}$ threshold of 100~meV/atom can be considered as possible electrodes. Thus, possible structures that can be considered as NIB electrodes, given thermodynamic stability constraints, include TiOF$_2$-NaTiOF$_2$, VOF$_2$-NaVOF$_2$, CrOF$_2$, and NaMnOF$_2$, and the ease of Na-ion mobility within these frameworks will further determine their suitability. Note that the high instabilities of NaCrOF$_2$ and MnOF$_2$ may limit the Na insertion/extraction capacity in these electrodes, compared to TiOF$_2$-NaTiOF$_2$, and VOF$_2$-NaVOF$_2$. Although we find NaTiO$_{2}$F to be metastable, we did not calculate Na E$_m$ within this structure given the negative average intercalation voltage associated with Na$_2$TiO$_2$F formation. 

\subsection{Ionic mobility}
\begin{figure*}
    \centering
    \includegraphics[width=\linewidth]{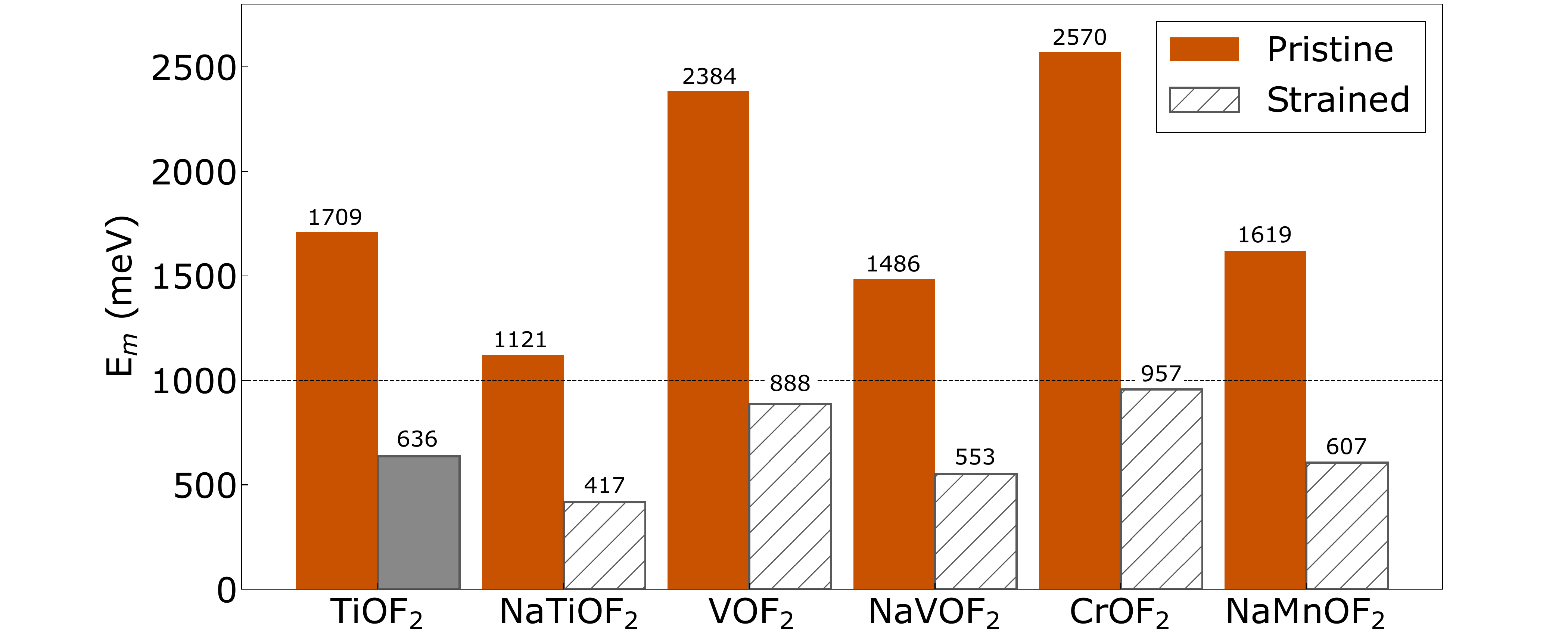}
    \caption{GGA-calculated E$_{m}$ of pristine (solid orange bars) and strained (solid grey or hashed bars) candidate compositions. The horizontal dotted line indicates the threshold E$_{m}$ of 1000~meV.}
    \label{fig:barrier}
\end{figure*}

For the candidate compositions identified via our voltage and stability calculations, we estimated the Na E$_m$ via the vacancy-mediated mechanism, and compiled the values in \textbf{Figure~{\ref{fig:barrier}}}. Barriers calculated in regular TMOF compositions are represented by solid orange bars in \textbf{Figure~{\ref{fig:barrier}}}, while barriers calculated in strained compositions (\textit{vide infra}) are indicated by solid grey or hatched bars. We used a threshold value of 1000~meV for the E$_m$, indicated by the dotted black line in \textbf{Figure~{\ref{fig:barrier}}}, to represent an electrode material that can be used in reasonable electrochemical conditions, similar to our previous works.\cite{tekliye2022exploration,lu2021searching} Thus, electrodes that exhibit E$_m \leq$1000~meV are considered candidates for further experimental exploration. Notably, all shortlisted TMOF compositions exhibit barriers that are above the 1000~meV threshold in their pristine state. Only NaTiOF$_{2}$, with a barrier of 1121~meV, is close to the 1000~meV threshold, with other compositions exhibiting significantly higher E$_m$, including NaVOF$_{2}$ (1486~meV), NaMnOF$_{2}$ (1619~meV), TiOF$_{2}$ (1709~meV), VOF$_{2}$ (2384~meV), and CrOF$_{2}$ (2570~meV). 

Introducing strain in an electrode can often lead to lowering of E$_m$ and consequent increasing in ionic mobility.\cite{wen2015interfacial,tealdi2016feeling,jia2017exploring} Thus, to examine whether the identified TMOFs can reasonably function as NIB electrodes under strain,  we introduced a homogeneous tensile strain of 5\% across all lattice parameters of TiOF$_{2}$ and evaluated the Na-E$_m$ using GGA-based NEB. To ensure that the tensile strain is maintained during structural relaxation, we restricted the relaxation of the end points to only include changes in ionic positions. Importantly, the strain introduction significantly reduced the E$_m$ to 636~meV (i.e., by 62.8\%) compared to the pristine-TiOF$_2$, as shown by the solid gray bar in \textbf{Figure~{\ref{fig:barrier}}}. 

Assuming similar reductions in calculated E$_m$ with strain addition in other TMOFs (i.e., by 62.8\% compared to the pristine-case), we estimate the barriers in strained NaTiOF$_2$, VOF$_2$, NaVOF$_2$, CrOF$_2$, and NaMnOF$_2$ to be 417~meV, 888~meV, 553~meV, 957~meV, and 607~meV, respectively (see hashed bars in \textbf{Figure~{\ref{fig:barrier}}}). Thus, all TMOFs identified using our voltage and stability criteria may exhibit reasonable Na-ionic mobility, provided a homogeneous strain is introduced within the materials. In practice, lattice expansion can be achieved through doping,\cite{liao2016mo,sun2020doping} heat/mechanical treatment,\cite{thomas2018atomistic,khan2023role} and/or epitaxially, such as in the case of thin film electrodes.\cite{tealdi2016feeling,yang2020strain} Note that introducing strain and maintaining it during electrochemical cycling may come at the cost of energy density of the eventual battery. Thus, we expect TiOF$_{2}$-NaTiOF$_{2}$, VOF$_{2}$-NaVOF$_{2}$, CrOF$_{2}$, and NaMnOF$_{2}$ to be NIB electrodes worth exploring experimentally, especially for strained electrode configurations and thin film batteries.

\section{Discussion}
In this work, we performed first principles calculations to explore the scope of 3$d$ TM-based F- and O-rich perovskite oxyfluorides (Na$_{x}$MOF$_{2}$ and Na$_{1+x}$MO$_2$F, $x$= 0-1) as NIB electrodes. Using a structural template based workflow, we identified the ground state polymorphs of the charged MOF$_2$ and NaMO$_2$F compositions (M = Ti, V, Cr, Mn, Fe, Co, or Ni) among six possible space groups commonly adopted by perovskites. Subsequently, we introduced Na to create corresponding discharged perovskite compositions, namely NaMOF$_2$ and Na$_2$MO$_2$F, and evaluted the average Na (de)intercalation voltages, 0~K thermodynamic stabilities in all perovskites, and Na-ion mobility in a select set of candidate perovskites. Based on our voltage, stability, and mobility calculations, we identify six perovskite compositions, namely TiOF$_2$-NaTiOF$_2$, VOF$_2$-NaVOF$_2$, CrOF$_2$, and NaMnOF$_2$ to hold some promise as NIB electrodes, if used in strained configurations and/or in thin film batteries. 

During the process of enumerating possible structures for the charged perovskites, we only considered a maximum of 16 lowest electrostatic energy configurations within each space group, and identified the ground state configuration among these structures as the one with the lowest DFT total energy (per f.u.). Note that the choice of the 16 lowest electrostatic energy structures (per space group) is an approximation and there is always a non-zero chance of encountering the `true' ground state beyond this choice. Using our criteria of a maximum of 16 structures per space group contributes to a total of 72 structures per perovskite composition (i.e., 16$\times 4+5+3$), which in turn adds up to 1008 structures over all TMs considered and over both O-rich and F-rich compositions, which by itself represents a significant computational expense.  Nevertheless, even if the `true' ground state is beyond the set of configurations we have considered here, we expect it to exhibit a lower energy, of the order of $\sim$10~meV/f.u., compared to the ground state that we have identified, which will only cause a marginal change to the voltage ($\sim$10~mV) and stability (E$^{hull} \pm$10~meV/f.u.) predictions.

Another approximation in our structure generation workflow is the identification of ground state configurations at the charged perovskite compositions followed by addition of Na to the lowest-energy charged structure to obtain the discharged structure. We could have followed a similar procedure of ground state identification using the discharged composition instead of the charged composition. Our choice of the charged perovskite composition for ground state identification was motivated largely by experimental reports on TMOFs in LIBs, wherein, Li was typically inserted into charged TMOF compositions.\cite{reddy2006metal,kuhn2020redox} Thus, the TMOF composition was synthesized first followed by Li discharge to obtain the discharged state. Considering the ground state configuration at the discharged state may lead to qualitatively different results, in terms of average voltages, 0~K stability, and Na-ionic mobility. But considering a workflow along the discharged compositions represents a significant computational effort, which we plan to take up as future work.

For all SCAN+$U$ calculations, we used the $U$ value derived from TM oxides since the oxide-based $U$ better reproduced the experimentally determined voltages for Li-intercalation in TMOFs, such as VO$_{2}$F $\leftrightarrow$ LiVO$_{2}$F and TiOF$_{2}$ $\leftrightarrow$ Li$_{0.5}$TiOF$_{2}$.\cite{reddy2006metal,kuhn2020redox} Hence, we did not tailor our $U$ values specifically for oxyfluorides. More experimental data will be needed to verify if such tailored $U$ values will yield more accurate predictions. Additionally, we initialised all our TMs in their corresponding ferromagnetic high-spin configurations, and did not consider possible magnetic/spin orderings due to their computational complexity, which may have marginally affected the set of ground states that we obtained. Also, SCAN+$U$ is known to overestimate intercalation voltages and meta/instability of compounds,\cite{long2021assessing} which is also a reason for us to consider a fairly large threshold (E$^{hull} \leq$ 100~meV/atom) for synthesizability. 

We used a reasonably high threshold for ionic mobility (E$_m \leq$1000~meV\cite{tekliye2022exploration}) to identify candidates partly due to the limited literature on Na-ion mobility within crystalline ordered oxyfluorides. Moreover, addition of F$^-$ to oxides can result in a reduction of Na$^+$ mobility due to more ionic Na-F bonds than Na-O, similar to observations of reduction in Li-mobility in F-doped oxide-based disordered rocksalts.\cite{ji2019hidden,ouyang2020effect} Importantly, our calculations indicated that only NaTiOF$_{2}$ (E$_{m} =$1121~meV) came close to the threshold used, with all other oxyfluorides considered exhibiting significantly high E$_m$ Na-motion. However, introducing a homogenous strain ($\sim$5\%) within the lattice can significantly reduce the E$_m$ (by $\sim$60\%), as demonstrated for the case of TiOF$_2$. Thus, TMOFs can exhibit reasonable rate performance under lattice strain. However, the need to maintain the strained structure may limit the applicability of TMOFs to low power and/or thin film batteries that are typically used in internet-of-things applications and wearable electronics.

Considering the oxyfluoride compositions of Na$_x$MOF$_2$ and Na$_{1+x}$MO$_2$F was primarily motivated by the availability of the M$^{4^{+}/3^{+}}$ redox couple, which is exhibited by several 3$d$ TMs, quite reversibly. Our work can be extended to other fluorine-added compositions, such as F-substituted oxides, phosphates, sulphates, and pyrophosphates. Indeed, high voltages and capacities with Na (de)intercalation have already been reported in fluorophosphates.\cite{zhu2017high,broux2016strong,kawai20214} Therefore, we are hopeful that our research lays the foundation for exploring other promising compositions for NIB electrodes within and beyond the chemical space of oxyfluorides.

\section{Conclusion}
NIBs, which represent an alternative technological pathway to the state-of-the-art LIBs in energy storage technology, require novel materials to improve the energy and power densities so that NIBs compete better with LIBs. Here, we explored the chemical space of perovskite-based TMOFs, considering both O-rich and F-rich compositions, as possible Na-ion intercalation hosts. Specifically, we performed DFT-based calculations on Na$_{x}$MOF$_{2}$ and Na$_{1+x}$MO$_{2}$F (x = 0-1), where M = Ti, V, Cr, Mn, Fe, Co, or Ni, evaluating the ground state polymorphs, average Na (de)intercalation voltages, 0~K stabilities, and Na$^+$ mobilities. We found that F-rich perovskites exhibit higher voltages than O-rich compositions, due to the stronger inductive effect of F$^-$. In terms of stability, only TiOF$_{2}$ and VOF$_{2}$ were stable while other compositions, including NaTiOF$_{2}$, NaVOF$_{2}$, CrOF$_{2}$ and NaMnOF$_{2}$ were metastable (E$^{hull} \leq$ 100~meV/atom). However, all stable and metastable TMOFs exhibited high E$_m$ ($\geq$1000~meV) for Na$^+$ motion in their pristine states. Nevertheless, introducing a 5\% homogenous tensile strain causes the E$_m$ to drop by $\sim$60\% compared to the pristine state, suggesting that the TMOFs may have applications in thin film batteries and in strained electrode configurations. Our study represents a systematic computational exploration of the oxyfluoride chemical space, which we hope will reinvigorate research in these chemistries for NIBs and beyond.

\section*{Acknowledgments}
G.S.G. acknowledges financial support from the Indian Institute of Science (IISc) and support from the Science and Engineering Research Board (SERB) of Government of India, under Sanction Numbers SRG/2021/000201 and IPA/2021/000007. D.D. thanks the Ministry of Human Resource Development, Government of India, for financial assistance. The authors acknowledge the computational resources provided by the Supercomputer Education and Research Centre (SERC), IISc. A portion of the calculations in this work used computational resources of the supercomputer Fugaku provided by RIKEN through the HPCI System Research Project (Project ID hp220393). We acknowledge National Supercomputing Mission (NSM) for providing computing resources of ‘PARAM Siddhi-AI’, under National PARAM Supercomputing Facility (NPSF), C-DAC, Pune, and the resources of ‘Param Utkarsh’ at CDAC Knowledge Park, Bengaluru. Both PARAM Siddhi-AI and PARAM Utkarsh are implemented by CDAC and supported by the Ministry of Electronics and Information Technology (MeitY) and Department of Science and Technology (DST), Government of India. The authors gratefully acknowledge the computing time provided to them on the high-performance computer noctua1 and noctua2 at the NHR Center PC2. This was funded by the Federal Ministry of Education and Research and the state governments participating on the basis of the resolutions of the GWK for the national high-performance computing at universities (www.nhr-verein.de/unsere-partner). The computations for this research were performed using computing resources under project hpc-prf-emdft.

\subsection*{Conflicts of Interest}
There are no conflicts to declare.

\section*{Supplementary Materials}
Electronic Supporting Information is available online at , Details of PAW potentials used, description of site determination for introducing additional Na in perovskites, compilation of 0~K phase diagrams and associated stability data, all calculated MEPs.

\section{Data and code availability}
The data that support the findings of this study are openly available at our \href{https://github.com/sai-mat-group/na-oxyfluorides/}{GitHub} repository.


\bibliographystyle{unsrt}
\bibliography{main}

\end{document}